\documentclass[         %
aps,                    
prd,                    
showpacs,               
nofootinbib,            
showkeys,               %
preprintnumbers,        %
floatfix]               
{revtex4}               
\usepackage{graphicx,longtable}

\begin{document}
\title{Supersymmetry and Nuclear Pairing}
\author{        A.~B. Balantekin}
\email{         baha@physics.wisc.edu}
\author{        Y. Pehlivan}
\email{         yamac@physics.wisc.edu}
\affiliation{  Department of Physics, University of Wisconsin\\
               Madison, Wisconsin 53706 USA }
\date{\today}
\begin{abstract}
We show that nuclear pairing Hamiltonian exhibits supersymmetry in the 
strong-coupling limit.  
The underlying supersymmetric quantum mechanical structure explains 
the degeneracies 
between the energies of the $N$ and $N_{\rm max} -N +1$ pair eigenstates.
The supersymmetry transformations connecting these states are given. 
\end{abstract}
\medskip
\preprint{} 
\maketitle


Supersymmetry concepts were utilized in several nuclear physics 
applications. Dynamical supersymmetries relate the spectra of even-even 
nuclei, considered as states of a system of correlated fermion pairs 
approximated as bosons, and odd-even nuclei, considered as states of a 
system of such bosons plus unpaired fermions 
\cite{BahaBalantekin:1981kt,Balantekin:1982bk}. 
Supersymmetric considerations are also used in various applications of 
random matrix theories \cite{Guhr:1997ve}. 
For example, invariance groups of the second moments arising in 
the study of compound nucleus scattering are orthosymplectic supergroups
\cite{Verbaarschot:1985jn}. Similar considerations arise in random 
matrix models mimicking QCD phase transitions \cite{Jackson:1995nf}.
Finally, it was shown that the spherical Nilsson 
Hamiltonian of the nuclear shell 
model has a dynamical Osp (1/2) supersymmetry connecting the SU(3) 
symmetry and pseudo SU(3) symmetry \cite{Balantekin:1992qp}. 

In this article we would like to discuss a fourth application of 
supersymmetry techniques to nuclear physics, namely to the spectra of 
nuclear pairing Hamiltonian:
\begin{equation}\label{1}
\hat{H}=\sum_{jm} \epsilon_j a^\dagger_{j\>m} a_{j\>m} -
|G|\sum_{jj'}c_{jj'} \hat{S}^+_j \hat{S}^-_{j'},
\end{equation}
where the pairing interaction between time-reversed states is written 
in terms of the quasi-spin operators
\begin{equation}\label{2}
\hat{S}^+_j=\sum_{m>0} (-1)^{(j-m)}
a^\dagger_{j\>m}a^\dagger_{j\>-m},\ \ \ \ \ \ \ \
\hat{S}^-_j=\sum_{m>0} (-1)^{(j-m)} a_{j\>-m}a_{j\>m}.\nonumber
\end{equation}
If the pairing interaction is separable, i.e. $c_{jj'}=c^*_jc_{j'}$, then 
in the strong coupling limit ($|G|\gg \epsilon_j$) the 
Hamiltonian of Eq. (\ref{1}) can be written as 
\begin{equation}\label{3}
\hat{H}_{SC} \sim -|G|\hat{S}^+(0)\hat{S}^-(0), 
\end{equation}
where we introduced the notation
\begin{equation}\label{4}
\hat{S}^+(0)=\sum_j c^*_j\hat{S}^+_j \ \ \ \ \mbox{and} \ \ \ \ \
\hat{S}^-(0)=\sum_j c_j\hat{S}^-_j .
\end{equation}

Supersymmetric quantum mechanics has been extensively studied 
\cite{Witten:1981nf,Cooper:1994eh,Fricke:1987ft}. Supersymmetric 
quantum mechanics relates the spectra of the Hamiltonians of the form 
$\hat{A}^{\dagger} \hat{A}$ and $\hat{A} \hat{A}^{\dagger}$, where $\hat{A}$ 
is an arbitrary operator. Clearly if one sets $\hat{A} = \hat{S}^-(0)$, then 
the separable Hamiltonian in 
Eq. (\ref{3}) and its supersymmetric partner have related spectra. However, 
in this note we would like to 
highlight a more subtle supersymmetry. To this end we first 
introduce the operator
\begin{equation}
\label{5}
\hat{T} = \exp \left( - i \pi \hat{S}^{(1)} \right)
\end{equation}
where $\hat{S}^{\pm,0} = \sum_j \hat{S}^{\pm,0}_j$. This operator exchanges 
$\hat{S}^+(0)$ and $\hat{S}^-(0)$
\begin{equation}
\label{6}
\hat{T}^\dagger \hat{S}^\pm(0) \hat{T} = \hat{S}^\mp (0). 
\end{equation}
Writing $\hat{S}^{(1)} = (1/2)( \hat{S}^+ + \hat{S}^-)$, it is easy to see 
that the operator $\hat{T}$ transforms the empty shell (i.e particle vacuum),
denoted by $|0\rangle$, to the fully occupied shell, denoted by
$|\bar{0}\rangle$:
\begin{equation}
\label{7}
\hat{T} |0\rangle=|\bar{0}\rangle .
\end{equation}
It is also straightforward to show that $\hat{T}$ 
converts states with $N$ particle pairs into states with
$N$ hole pairs. 

To exhibit the supersymmetry of the pairing Hamiltonian we next define the 
operators
\begin{equation}\label{8}
\hat{B}^- = \hat{T}^{\dagger} \hat{S}^-(0),\ \ \ \ \ \ \ \
\hat{B}^+ = \hat{S}^+(0) \hat{T}. 
\end{equation}
Supersymmetric quantum mechanics tells us that the partner Hamiltonians 
$\hat{H}_1 = \hat{B}^+ \hat{B}^-$ and $\hat{H}_2 = \hat{B}^- \hat{B}^+$ 
have identical spectra except 
for the ground state of $\hat{H}_1$. This ground state has zero energy (i.e., 
it is annihilated by  $\hat{B}^-$, or alternately by  $\hat{S}^-(0)$). 
The states of the $\hat{H}_1 = \hat{B}^+ \hat{B}^-$, $| \Psi_1 
\rangle$, 
and $\hat{H}_2 = \hat{B}^- \hat{B}^+$, $|\Psi_2 \rangle$, are related 
\begin{equation}
\label{8a}
| \Psi_2 \rangle \sim \hat{B}^- | \Psi_1 \rangle = 
\hat{T}^{\dagger} \hat{S}^-(0) | \Psi_1 \rangle 
=  \hat{S}^+(0) \hat{T}^{\dagger} | \Psi_1 \rangle .  
\end{equation}
Using Eq. (\ref{6}), one can easily show that the two Hamiltonians 
$\hat{H}_1$ and $\hat{H}_2$ are 
actually identical and equal to the pairing Hamiltonian in the 
strong-coupling limit:
\begin{equation}
\label{9}
\hat{B}^+ \hat{B}^- = \hat{B}^- \hat{B}^+ =  \hat{S}^+(0)\hat{S}^-(0), 
\end{equation}
hence the role of the supersymmetry is to connect the states 
$| \Psi_2 \rangle$ and $| \Psi_1 \rangle$. 
Clearly even though these two Hamiltonians are identical, the corresponding 
eigenstates are not. For example the particle vacuum, $|0\rangle$, is 
annihilated by $\hat{B}^-$, but not by $\hat{B}^+$.  
Below we show that 
$| \Psi_1 \rangle$ represent the particle states and 
$| \Psi_2 \rangle$ represent the hole states. Hence supersymmetry 
connects particle and hole states. 

The energy eigenvalues and eigenstates of the Hamiltonian in Eq. (\ref{3}) 
are worked out in detail in Refs. 
\cite{Pan:1997rw} and \cite{Balantekin:2007vs} and sketched in the Appendix. 

Let us first consider one-pair 
states. The state in Eq. ({\ref{a2}) of the Appendix is annihilated 
by $\hat{B}^-$, hence it has no supersymmetric partner. Using Eq. (\ref{8a}), 
we can find the supersymmetric partner of the non-zero energy state given in 
Eq. (\ref{a4}). We obtain
\begin{equation}
\label{10}
\hat{B}^- \hat{S}^+(0)|0\rangle = \hat{T}^{\dagger} 
[\hat{S}^-(0),\hat{S}^+(0)]|0\rangle. 
\end{equation}
Since the action of the commutator on the particle vacuum gives a number we 
find 
\begin{equation}
\label{10a}
\hat{B}^- \hat{S}^+(0)|0\rangle \sim  \hat{T}^{\dagger} |0\rangle =
 |\bar{0}\rangle ,
\end{equation}
i.e., the supersymmetric partner of the one-pair state is the completely 
filled state (state with $N_{\rm max}$ number of pairs). 
In general the supersymmetric partners of N-pair states are states with 
$N_{\rm max} -N +1$ pairs. 
To see this consider the state given in Eq. (\ref{a10}) with $(N-1)$ hole 
pairs (or equivalently with $(N_{\rm max}-N+1)$ particle pairs). Its 
supersymmetric partner should be given by  
\begin{equation}
\label{11}
\hat{B}^+ 
\hat{S}^-(z_1^{(N)})\hat{S}^-(z_2^{(N)})\dots\hat{S}^-(z_{N-1}^{(N)}) 
|\bar{0}\rangle = 
\hat{S}^+(0) \hat{T}
\hat{S}^-(z_1^{(N)}) \hat{T}^{\dagger} \hat{T}
\hat{S}^-(z_2^{(N)}) \hat{T}^{\dagger} \dots \hat{T}
\hat{S}^-(z_{N-1}^{(N)}) \hat{T}^{\dagger} \hat{T} 
|\bar{0}\rangle ,
\end{equation}
which, using Eqs. (\ref{6}) and (\ref{7}), reduces to 
\begin{equation}
\label{12}
\hat{S}^+(0)\hat{S}^+(z^{(N)}_1) \dots
\hat{S}^+(z^{(N)}_{N-1})|0\rangle . 
\end{equation}
This is nothing but the state with N-pairs given in Eq. (\ref{a7}) of the 
Appendix. Hence, for $N \le N_{\rm max} /2$, the states with $N$ pairs and 
$(N_{\rm max}-N+1)$ pairs have the same energy as was shown in Ref. 
\cite{Balantekin:2007vs} using an entirely different approach. The  
resulting supersymmetric spectra of the nuclear pairing 
Hamiltonian are illustrated in Fig. 1. Clearly this 
supersymmetry is broken by the 
single-particle energies (i.e. the Hamiltonian in Eq. (\ref{3}) exhibits 
this supersymmetry while the Hamiltonian in Eq. (\ref{1}) does not). 

\begin{figure}[h] \begin{center}
\vspace*{2cm}
\includegraphics[scale=0.6]{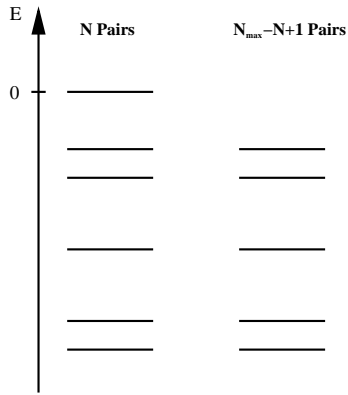}
\vspace*{+0.5cm} \caption{ \label{fig:1}
Spectra of nuclear pairing Hamiltonian exhibiting supersymmetry.
States with $N$ pairs and with  
$(N_{\rm max}-N+1)$ pairs are supersymmetric partners of each other 
($N \le N_{\rm max} /2)$. }
\end{center}  \end{figure}

We showed that nuclear pairing Hamiltonian exhibits supersymmetry in the 
strong-coupling limit. 
Provided that $N \le N_{\rm max} /2$, the states with $N$ pairs and 
$(N_{\rm max}-N+1)$ pairs are supersymmetric partners of each other. 
The underlying supersymmetric quantum mechanical structure explains 
i) Existence of the zero-energy states when the number of pairs are less 
than or equal to the half of the maximum allowed value, 
ii) Degeneracies 
between the energies of the $N$ and $N_{\rm max} -N +1$ pair eigenstates.

\section*{ACKNOWLEDGMENTS}
This work was supported in part by the U.S. National Science
Foundation Grant No.\ PHY-0555231 and in part by 
the University of Wisconsin Research Committee with funds granted by
the Wisconsin Alumni Research Foundation. 

\section*{Appendix}

Here we summarize the information about the eigenvalues and eigenstates 
of the Hamiltonian of Eq. (\ref{3}). Full details can be found in Refs. 
\cite{Pan:1997rw} and \cite{Balantekin:2007vs}. 
One first introduces the operators
\begin{equation}\label{a1}
\hat{S}^+(x)=\sum_j\frac{c^*_j}{1-|c_j|^2x}\hat{S}^+_j \ \ \ \ \
\mbox{and} \ \ \ \ \
\hat{S}^-(x)=\sum_j\frac{c_j}{1-|c_j|^2x}\hat{S}^-_j, 
\end{equation}
where $x$ is, in general, a complex parameter. Clearly the operators given  
in Eq.(\ref{4}) are these operators with $x=0$. For a single pair there are 
two kinds of states. The state
\begin{equation}\label{a2}
\hat{S}^+(x^{(1)})|0\rangle 
\end{equation}
is an eigenstate of the Hamiltonian of Eq. (\ref{3}) with zero energy, 
if $x^{(1)}$ 
satisfies the Bethe ansatz equation 
\begin{equation}\label{a3}
\sum_j \frac{-\Omega_j/2}{1/|c_j|^2-x^{(1)}}=0. 
\end{equation}
There are as many states with zero energy as different solutions of Eq. 
(\ref{a3}).  
In addition the state 
\begin{equation}\label{a4}
\hat{S}^+(0)|0\rangle
\end{equation}
is an eigenstate with energy 
\begin{equation}
\label{a4a}
E = - |G| \sum_j \Omega_j |c_j|^2.
\end{equation}
If the 
available orbitals are less than half full 
(or at most half full) these results 
generalize in the following manner: 
The state 
\begin{equation} \label{a5}
\hat{S}^+(x^{(N)}_1)\hat{S}^+(x^{(N)}_2) \dots
\hat{S}^+(x^{(N)}_N)|0\rangle 
\end{equation}
is an eigenstate with zero energy if the Bethe ansatz equations
\begin{equation}
\label{a6}
 \sum_j
\frac{-\Omega_j/2}{1/|c|_j^2-x^{(N)}_m}=\sum_{k=1(k\neq m)}^N
\frac{1}{x^{(N)}_m-x^{(N)}_k} 
\end{equation}
are satisfied. 
In addition the state
\begin{equation}
\label{a7}
\hat{S}^+(0)\hat{S}^+(z^{(N)}_1) \dots
\hat{S}^+(z^{(N)}_{N-1})|0\rangle 
\end{equation}
is an eigenstate with energy
\begin{equation}
\label{a8}
E = - |G| \left(
 \sum_j
\Omega_j |c_j|^2-\sum_{k=1}^{N-1} \frac{2}{z^{(N)}_k} \right), 
\end{equation}
if the following Bethe ansatz equations are satisfied: 
\begin{equation}
\label{a9}
\sum_j \frac{-\Omega_j/2}{1/|c_j|^2-z^{(N)}_m}
=\frac{1}{z^{(N)}_m}+\sum_{k=1(k\neq m)}^{N-1}
\frac{1}{z^{(N)}_m-z^{(N)}_k}.
\end{equation}
When the orbitals are more than half full there are no zero energy states. 
The completely filled state $|\bar{0}\rangle$ has an energy given by 
Eq. (\ref{a4a}). In addition the state 
\begin{equation}
\label{a10}
\hat{S}^-(z_1^{(N)})\hat{S}^-(z_2^{(N)})\dots\hat{S}^-(z_{N-1}^{(N)}) 
|\bar{0}\rangle
\end{equation}
has the same energy as in Eq. (\ref{a8}) with the parameters $z^N_i$ 
satisfying the Bethe ansatz equations of Eq. (\ref{a9}). 
It should be noted that, although the pairing problem outlined above 
is exactly solvable, it is not easy to find solutions of the resulting 
non-linear Bethe ansatz equations. Several strategies to solve these equations 
are discussed in Refs. \cite{Dukelsky:2001dd} and 
\cite{Balantekin:2004yf}.


\end{document}